\documentclass[aps,prd,twocolumn,superscriptaddress]{revtex4}
\usepackage{epsfig,epsf}
\usepackage{amsmath}
\usepackage{amsthm}
\usepackage{amsfonts}
\usepackage{amssymb}
\usepackage{dsfont}
\usepackage{multirow}
\usepackage{appendix}
\usepackage{slashed}
\usepackage[active]{srcltx}
\usepackage{psfrag}
\newcommand{\func}[1]{\operatorname{#1}}
\begin{document}

\title{Semileptonic decays of the scalar tetraquark $Z_{bc;\overline{u}%
\overline{d}}^{0}$}
\date{\today}
\author{H.~Sundu}
\affiliation{Department of Physics, Kocaeli University, 41380 Izmit, Turkey}
\author{S.~S.~Agaev}
\affiliation{Institute for Physical Problems, Baku State University, Az--1148 Baku,
Azerbaijan}
\author{K.~Azizi}
\thanks{Corresponding author}
\affiliation{Department of Physics, University of Tehran, North Karegar Ave., Tehran
14395-547, Iran}
\affiliation{Department of Physics, Do\v{g}u\c{s} University, Acibadem-Kadik\"{o}y, 34722
Istanbul, Turkey}
\affiliation{School of Particles and Accelerators, Institute for Research in Fundamental
Sciences (IPM) P.O. Box 19395-5531, Tehran, Iran}

\begin{abstract}
We study semileptonic decays of the scalar tetraquark $Z_{bc;\overline{u}%
\overline{d}}^{0}$ to final states $T_{bs;\overline{u}\overline{d}%
}^{-}e^{+}\nu_{e}$ and $T_{bs;\overline{u}\overline{d}}^{-}\mu^{+}\nu_{\mu}$%
, which run through the weak transitions $c\to se^{+}\nu_{e}$ and $c\to
s\mu^{+}\nu_{\mu}$, respectively. To this end, we calculate the mass and
coupling of the final-state scalar tetraquark $T_{bs;\overline{u}\overline{d}%
}^{-} $ by means of the QCD two-point sum rule method: these spectroscopic
parameters are used in our following investigations. In calculations we take
into account the vacuum expectation values of the quark, gluon, and mixed
operators up to dimension ten. We use also three-point sum rules to evaluate
the weak form factors $G_{i}(q^2)$ ($i=1,~2$) that describe these decays.
The sum rule predictions for $G_{i}(q^2)$ are employed to construct fit
functions $F_{i}(q^2)$, which allow us to extrapolate the form factors to
the whole region of kinematically accessible $q^2$. These functions are
required to get partial widths of the semileptonic decays $\Gamma \left(
Z_{bc}^{0}\rightarrow Te^{+}\nu_{e}\right) $ and $\Gamma \left(
Z_{bc}^{0}\rightarrow T\mu ^{+}\nu_{\mu }\right)$ by integrating
corresponding differential rates. We analyze also the two-body nonleptonic
decays $Z_{bc;\overline{u}\overline{d}}^{0} \to T_{bs;\overline{u}\overline{d%
}}^{-}\pi^{+}$ and $Z_{bc;\overline{u}\overline{d}}^{0} \to T_{bs;\overline{u%
}\overline{d}}^{-}K^{+}$, which are necessary to evaluate the full width of
the $Z_{bc;\overline{u}\overline{d}}^{0}$. The obtained results for $\Gamma_{%
\mathrm{full}}=(3.18\pm 0.39)\times 10^{-11}~\mathrm{MeV}$ and mean lifetime
$20.7_{-2.3}^{+2.9}~\mathrm{ps}$ of the tetraquark $Z_{bc;\overline{u}%
\overline{d}}^{0}$ can be used in experimental investigations of this exotic
state.
\end{abstract}

\maketitle


\section{Introduction}

\label{sec:Int}
Investigations of double-heavy tetraquarks composed of a heavy $QQ$ diquark [%
$Q$ is the heavy $c$ or $b$ quark] and a light antidiquark are among
interesting topics in physics of exotic hadrons. The interest to such kind
of quark configurations is connected with a possible stability of some of
them against the strong and electromagnetic decays. The relevant problems
were addressed already in the pioneering papers \cite%
{Ader:1981db,Lipkin:1986dw,Zouzou:1986qh}, in which a stability of the
exotic four-quark mesons $QQ\bar{Q}\bar{Q}$ and $QQ\bar{q}\bar{q}$ was
examined. It was found that the heavy $Q$ and light $q$ quarks with a large
mass ratio $m_{Q}/m_{q}$ may form the stable tetraquarks $QQ\bar{q}\bar{q}$.
The similar conclusions were drawn in Ref.\ \cite{Carlson:1987hh} as well,
in accordance of which the isoscalar $J^{P}=1^{+}$ tetraquark $T_{bb;%
\overline{u}\overline{d}}^{-}$ lies below the two B-meson threshold and can
decay only weakly.

All available theoretical tools of high energy physics were exploited to
study properties of double-heavy exotic mesons; the chiral and dynamical
quark models, the relativistic quark model and sum rules method were
mobilized to calculate their parameters \cite%
{Pepin:1996id,Janc:2004qn,Cui:2006mp,Vijande:2006jf,Ebert:2007rn,Navarra:2007yw,Du:2012wp,Hyodo:2012pm,Esposito:2013fma}%
. Interest to these mesons renewed after experimental observation by the
LHCb Collaboration of the $\Xi _{cc}^{++}=ccu$ baryon \cite{Aaij:2017ueg}.
Its mass was used as an input information in a phenomenological model to
estimate the mass of the axial-vector tetraquark $T_{bb;\overline{u}%
\overline{d}}^{-}$ \cite{Karliner:2017qjm}. The obtained prediction $%
m=(10389\pm 12)~\mathrm{MeV}$ is $215~\mathrm{MeV}$ below the $B^{-}%
\overline{B}^{\ast 0}$ threshold and $170~\mathrm{MeV}$ below the threshold
for decay $B^{-}\overline{B}^{0}\gamma $, which means that $T_{bb;\overline{u%
}\overline{d}}^{-}$ is stable against the strong and electromagnetic decays
and dissociates only weakly. The conclusion about the strong-interaction
stability of the tetraquarks $T_{bb;\overline{u}\overline{d}}^{-}$, $T_{bb;%
\overline{u}\overline{s}}^{-}$, and $T_{bb;\overline{d}\overline{s}}^{0}$
was made in Ref.\ \cite{Eichten:2017ffp} on the basis of the relations
derived from heavy-quark symmetry. The mass $m=10482~\mathrm{MeV}$ of the
axial-vector tetraquark $T_{bb;\overline{u}\overline{d}}^{-}$ found there is
$121~\mathrm{MeV}$ below the open-bottom threshold.

In Ref.\ \cite{Agaev:2018khe} we calculated the spectroscopic parameters of
the axial-vector tetraquark $T_{bb;\overline{u}\overline{d}}^{-}$ and
analyzed also its semileptonic decay to the scalar state $Z_{bc;\overline{u}%
\overline{d}}^{0}$. Our result for its mass $m=(10035~\pm 260)~\mathrm{MeV}$
confirms once more that it is stable against the strong and electromagnetic
decays. We evaluated the total width and mean lifetime of $T_{bb;\bar{u}\bar{%
d}}^{-}$ \ using the semileptonic decay channels $T_{bb;\overline{u}%
\overline{d}}^{-}\rightarrow Z_{bc;\overline{u}\overline{d}}^{0}l\bar{\nu
_{l}}$, where $l=e,\mu $ and $\tau $. The predictions $\Gamma =(7.17\pm
1.23)\times 10^{-8}~\mathrm{MeV}$ and $\tau =9.18_{-1.34}^{+1.90}~\mathrm{fs}
$ provide information useful for experimental investigation of the
double-heavy exotic mesons. Details of performed analysis and references to
earlier and recent articles devoted to different aspects of the doubly and
fully heavy tetraquarks can be found in Ref.\ \cite{Agaev:2018khe}.

We determined the mass and coupling of the scalar four-quark meson $Z_{bc;%
\overline{u}\overline{d}}^{0}$ (hereafter $Z_{bc}^{0}$) as well \cite%
{Agaev:2018khe}, because these parameters were necessary to evaluate the
width of the semileptonic decay $T_{bb;\overline{u}\overline{d}%
}^{-}\rightarrow Z_{bc}^{0}l\bar{\nu _{l}}$. For these purposes we employed
the QCD sum rule approach and found $m_{Z}=(6660\pm 150)~\mathrm{MeV}$. This
prediction is considerably below the threshold $7145~\mathrm{MeV}$ for
strong decays of $Z_{bc}^{0}$ to heavy mesons $B^{-}D^{+}$ and $\overline{%
B^{0}}D^{0}$. The state $Z_{bc}^{0}$ cannot decay to a pair of heavy and
light mesons as well; this fact differs it qualitatively from the open
charm-bottom scalar tetraquarks $cq\overline{b}\overline{q}$ and $cs%
\overline{b}\overline{s}$, which decay to $B_{c}\pi $ and $B_{c}\eta $
mesons \cite{Agaev:2016dsg}, respectively. The thresholds for the
electromagnetic decays $Z_{bc}^{0}\rightarrow \overline{B^{0}}%
D_{1}^{0}\gamma $ and $B^{\ast }D_{0}^{\ast }\gamma $ exceed $7600~\mathrm{%
MeV}$ and are higher than the mass of $Z_{bc}^{0}$. In other words, the
tetraquark $Z_{bc}^{0}$ as the state $T_{bb;\bar{u}\bar{d}}^{-}$ is the
strong- and electromagnetic-interaction stable particle.

The scalar and axial-vector states $bc\overline{u}\overline{d}$ were
subjects of interesting theoretical investigations with, sometimes,
controversial predictions. In fact, the analysis performed in Ref.\ \cite%
{Karliner:2017qjm} showed that $Z_{bc}^{0}$ resides $11~\mathrm{MeV}$ below
the threshold $7145~\mathrm{MeV}$ for $S$-wave decays to conventional heavy $%
B^{-}D^{+}$ and $\overline{B^{0}}D^{0}$ mesons. Computations of the
ground-state $QQ^{\prime }\overline{u}\overline{d}$ tetraquarks' masses
carried out in the context of the Bethe-Salpeter method led to similar
conclusions \cite{Feng:2013kea}. The mass of $Z_{bc}^{0}$ found there (for
some set of used parameters) equals to $6.93~\mathrm{GeV}$ and is lower than
the relevant strong threshold. On the contrary, for the masses of the scalar
and axial-vector $bc\overline{u}\overline{d}$ states the heavy-quark
symmetry predicts $7229~\mathrm{MeV}$ and $7272~\mathrm{MeV} $ \cite%
{Eichten:2017ffp}, which means that they can decay to ordinary mesons $%
B^{-}D^{+}/\overline{B^{0}}D^{0}$ and $B^{\ast }D$, respectively. The
charged exotic scalar mesons $Z_{bc;\overline{u}\overline{u}}^{-}$ and $%
Z_{bc;\overline{d}\overline{d}}^{+}$ were explored by means of the QCD sum
rule method as well \cite{Chen:2013aba}; the mass of these particles $%
m=(7.14\pm 0.10)~\mathrm{GeV}$ is higher than our prediction for $m_{Z}$.

The recent lattice simulations prove the strong-interaction stability of the
$I(J^{P})=0(1^{+})$ four-quark meson $Z_{ud;\overline{c}\overline{b}}^{0}$
with the mass in the range $15$ to $61$ $\mathrm{MeV}$ below $\overline{D}%
B^{\ast }$ threshold \cite{Francis:2018jyb}. But, because of theoretical
uncertainties the authors could not determine whether this tetraquark would
decay electromagnetically to $\overline{D}B\gamma $ or can transform only
weakly. Another confirmation of the $bc\overline{u}\overline{d}$ tetraquarks
stability came from Ref.\ \cite{Caramees:2018oue}; there it was demonstrated
that both the $J^{P}=0^{+}$ and $1^{+}$ isoscalar tetraquarks $bc\overline{u}%
\overline{d}$ are stable against the strong decays. The isoscalar $%
J^{P}=0^{+}$ state is also electromagnetic-interaction stable, whereas $%
J^{P}=1^{+}$ may undergo the electromagnetic decay to $\overline{B}D\gamma $.

In light all of these theoretical predictions, it becomes evident that
decays of the tetraquark $Z_{bc}^{0}$ are sources of a valuable information
about this exotic meson. In the present work we explore the semileptonic
decays of the tetraquark $Z_{bc}^{0}$ which are important for some reasons.
First of all, $Z_{bc}^{0}$ may be produced copiously at the LHC \cite%
{Ali:2018xfq}, hence it is necessary to fix processes, where it has to be
searched for. The second reason is exploration of the tetraquark $T_{bb;%
\overline{u}\overline{d}}^{-}$ itself, and decay channels appropriate for
its discovery. As usual, all states classified till now as candidates to
tetraquarks were seen through their decays to conventional mesons. If a
tetraquark is stable against strong and electromagnetic decays, then it
should be observed due to products of its weak decays. In the case under
discussion at the first stage $T_{bb;\overline{u}\overline{d}}^{-}$ decays
to $Z_{bc}^{0}$ and $l\bar{\nu _{l}}$. But, because the scalar tetraquark $%
Z_{bc}^{0}$ does not transform directly to conventional mesons, one needs to
consider its weak decays, as well.

The weak decays of $Z_{bc}^{0}$ can proceed through different channels. The
dominant semileptonic decay modes of $Z_{bc}^{0}$ are the processes $%
Z_{bc}^{0}\rightarrow T_{bs;\overline{u}\overline{d}}^{-}e^{+}\nu _{e}$ and $%
Z_{bc}^{0}\rightarrow T_{bs;\overline{u}\overline{d}}^{-}\mu ^{+}\nu _{\mu }$%
, which run due to transitions $c\rightarrow W^{+}s$ and $W^{+}\rightarrow
\overline{l}\nu _{l}$. The channels triggered by the decays $c\rightarrow
W^{+}d$ and $W^{+}\rightarrow \overline{l}\nu _{l}$ lead to creation of the
tetraquark $T_{bd;\overline{u}\overline{d}}^{-}$, and are suppressed
relative to the first modes by a factor $|V_{cd}|^{2}/|V_{cs}|^{2}\simeq 0.05
$. The similar arguments can be applied to other semileptonic decays of $%
Z_{bc}^{0}$ generated by a chain of transitions $b\rightarrow W^{-}c$ $%
\rightarrow cl\overline{\nu }_{l}$ and $b\rightarrow W^{-}u$ $\rightarrow ul%
\overline{\nu }_{l}$, respectively. In fact, the Cabibbo-Kobayashi-Maskawa
(CKM) matrix element $|V_{bc}|$, which is small numerically, and the ratio $%
|V_{bu}|^{2}/|V_{bc}|^{2}$ $\simeq 0.01$ demonstrates a subdominant nature
of the decays $b\rightarrow cl\overline{\nu }_{l}$ and $b\rightarrow ul%
\overline{\nu }_{l}$. The weak decay $c\rightarrow W^{+}s$ may be followed
by transitions $W^{+}\rightarrow u\overline{d}$ and $W^{+}\rightarrow u%
\overline{s}$, which give rise to nonleptonic decays of $Z_{bc}^{0}$. In the
hard-scattering mechanism, for example, a pair $u\overline{d}$ may form
ordinary mesons with $q\overline{q}$ quarks appeared due to a gluon from one
of $u$ or $\overline{d}$ quarks. These processes lead to final states $%
Z_{bc}^{0}\rightarrow T_{bs;\overline{u}\overline{d}}^{-}M_{1}(u\overline{q}%
)M_{2}(q\overline{d})$ which are suppressed relative to the semileptonic
decays by the factor $\alpha _{s}^{2}|V_{ud}|^{2}$. But $u\overline{d}$ and $%
u\overline{s}$ quarks can form $\pi ^{+}$ and $K^{+\text{ }}$mesons and
generate the two-body nonleptonic decays of the tetraquark $Z_{bc}^{0}$,
i.e., the processes $Z_{bc}^{0}\rightarrow T_{bs;\overline{u}\overline{d}%
}^{-}\pi ^{+}$ and $Z_{bc}^{0}\rightarrow T_{bs;\overline{u}\overline{d}%
}^{-}K^{+}$ . There is also a class of multimeson processes, when $u%
\overline{d}$ and $u%
\overline{s}$ combine directly with quarks from $T_{bs;\overline{u}%
\overline{d}}^{-}$ and create three-meson final states. The two-body and
three-meson nonleptonic decays do not suppressed by additional factors
relative to the semileptonic decays, and their contributions to full width
of $Z_{bc}^{0}$ may be sizeable. Parameters of these channels may provide a
valuable new information on features of the exotic meson $Z_{bc}^{0}$.

The tetraquark $T_{bs;\overline{u}\overline{d}}^{-}$ can bear different
quantum numbers. We treat $T_{bs;\overline{u}\overline{d}}^{-}$ as a scalar
particle, and in what follows denote it by $T$ . To calculate the width of
aforementioned decays, one needs the mass and coupling of the tetraquark $T$
; they enter as parameters to the sum rules for the weak form factors that
determine width of the decays. The spectroscopic parameters of this
tetraquark can be extracted from the two-point correlation function by means
of the sum rule approach, which is one of the powerful nonperturbative tools
in QCD \cite{Shifman:1978bx,Shifman:1978by}. It can be applied to compute
spectroscopic parameters and decay width not only of the conventional
hadrons but also the exotic states [for the recent review, see Ref.\ \cite%
{Albuquerque:2018jkn}].

In the present work the mass and coupling of $T$ are calculated by taking
into account vacuum expectation values of various quark, gluon, and mixed
local operators up to dimension ten. The weak form factors $G_{i}(q^{2})$, ($%
\,i=1,2$) are extracted from the QCD\ three-point sum rules, which allow us
to find numerical values of $G_{i}(q^{2})$ at momentum transfer $q^{2}$
accessible for sum rule computations. Later we fit $G_{i}(q^{2})$ by
functions $F_{i}(q^{2})$, and extrapolate them to a whole domain of physical
$q^{2}$. The fit functions are used to integrate the differential decay
rates and obtain the width of the semileptonic decays $\Gamma \left(
Z_{bc}^{0}\rightarrow Te^{+}\nu _{e}\right) $ and $\Gamma \left(
Z_{bc}^{0}\rightarrow T\mu ^{+}\nu _{\mu }\right) $. We also calculate the
widths of the nonleptonic decays $Z_{bc}^{0}\rightarrow T\pi ^{+}$ and $%
Z_{bc}^{0}\rightarrow TK^{+}$, and use this information to evaluate the full
width of $Z_{bc}^{0}$.

This article is structured in the following form: In Sec.\ \ref{sec:Masses}
we derive the QCD two-point sum rules for the mass and coupling of the
tetraquark $T$, and find their numerical values. In Sec. \ref{sec:Decays}
the QCD three-point correlation functions are utilized to get sum rules for
the form factors $G_{i}(q^{2})$. Here we carry out also numerical analysis
of derived expressions and determine the fit functions, and evaluate the
width of the semileptonic decays of concern. Section \ref{sec:Decays2} is
devoted to analysis of the two-body nonleptonic decays of the tetraquark $%
Z_{bc}^{0}$, where we calculate the partial widths of the processes $%
Z_{bc}^{0}\rightarrow T\pi ^{+}$ and $Z_{bc}^{0}\rightarrow TK^{+}$. In
Sec.\ \ref{sec:Analysis} we evaluate the full width and mean lifetime of $%
Z_{bc}^{0}$, and analyze decay channels of the tetraquarks $Z_{bc}^{0}$\ and
$T_{bb;\overline{u}\overline{d}}^{-}$. This section contains also our
concluding remarks.


\section{Spectroscopic parameters of the tetraquark $T_{bs;\overline{u}%
\overline{d}}^{-}$}

\label{sec:Masses}
The spectroscopic parameters of the tetraquark $T$ are important to
calculate the width of the exotic $Z_{bc}^{0}$ meson's semileptonic decays.
The $T$ state contains four quarks $b,s,u,$ and $d$ of different flavors and
has the heavy-light structure. In other words, the $b$-quark and $s$-quark,
which is considerably heavier than $q=u,d$ , groups to form the heavy
diquark, whereas the antidiquark is built of light $u$ and $d$ quarks. This
is the main difference of $T$ and the famous resonance $X(5568)$; the latter
has the same quark content, but $b$ and $s$ quarks are distributed between a
diquark and an antidiquark \cite{Agaev:2016mjb}. The scalar tetraquark $T$
can be composed using diquarks of a different type. The ground-state scalar
particle $T$ should be composed of the scalar diquark $\epsilon
^{abc}[b_{b}^{T}C\gamma _{5}s_{c}]$ in the color antitriplet and flavor
antisymmetric state and the antidiquark $\epsilon ^{ade}[\overline{u}%
_{d}\gamma _{5}C\overline{d}_{e}^{T}]$ in the color triplet state. The
reason is that they are most attractive diquark configurations, and exotic
mesons composed of them should be lighter and more stable than four-quark
mesons made of other diquarks \cite{Jaffe:2004ph}. Therefore, we assume that
$T$ has such favorable structure, and accordingly choose the interpolating
current $J(x)$
\begin{equation}
J(x)=\epsilon \widetilde{\epsilon }[b_{b}^{T}(x)C\gamma _{5}s_{c}(x)][%
\overline{u}_{d}(x)\gamma _{5}C\overline{d}_{e}^{T}(x)],  \label{eq:Curr1}
\end{equation}%
where $\epsilon \widetilde{\epsilon }=\epsilon ^{abc}\epsilon ^{ade}$. In
this expression $a,b,c,d$ and $e$ are color indices and $C$ is the
charge-conjugation operator.

The mass and coupling of the tetraquark $T$ can be obtained from the QCD
two-point sum rules. To derive the sum rules for the mass $m_{T}$ and
coupling $f_{T}$ of $T$, we analyze the correlation function
\begin{equation}
\Pi (p)=i\int d^{4}xe^{ipx}\langle 0|\mathcal{T}\{J(x)J^{\dag
}(0)\}|0\rangle .  \label{eq:CF1}
\end{equation}

To find the phenomenological side of the sum rule $\Pi ^{\mathrm{Phys}}(p)$,
we treat $T$ as a ground-state particle and use the "ground-state +
continuum" scheme. Then $\Pi ^{\mathrm{Phys}}(p)$ contains a contribution of
the ground-state particle and contributions arising from higher resonances
and continuum states
\begin{equation}
\Pi ^{\mathrm{Phys}}(p)=\frac{\langle 0|J|T(p)\rangle \langle
T(p)|J^{\dagger }|0\rangle }{m_{T}^{2}-p^{2}}+\ldots ,  \label{eq:Phys1}
\end{equation}%
which are denoted in Eq.\ (\ref{eq:Phys1}) by dots. This expression for the
phenomenological side is obtained by inserting into the correlation function
$\Pi (p)$ a full set of relevant states and carrying out integration in Eq.\
(\ref{eq:CF1}) over $x$.

Computation of $\Pi ^{\mathrm{Phys}}(p)$ can be continued by introducing the
matrix element of the scalar tetraquark
\begin{equation}
\langle 0|J|T(p)\rangle =f_{T}m_{T}.  \label{eq:ME1}
\end{equation}%
After simple manipulations we get%
\begin{equation}
\Pi ^{\mathrm{Phys}}(p)=\frac{f_{T}^{2}m_{T}^{2}}{m_{T}^{2}-p^{2}}+\ldots
\label{eq:Phys1a}
\end{equation}%
At the next step one should choose in $\Pi ^{\mathrm{Phys}}(p)$ some Lorentz
structure and fix the corresponding invariant amplitude. The correlation
function $\Pi ^{\mathrm{Phys}}(p)$ contains only the trivial structure $\sim
I$, therefore the amplitude $\Pi ^{\mathrm{Phys}}(p^{2})$ is given by the
function from Eq.\ (\ref{eq:Phys1a}).

We need also to determine $\Pi (p)$ by employing the perturbative QCD and
express it in terms of the quark propagators. For these purposes, we utilize
the explicit expression of the interpolating current $J(x)$ and calculate $%
\Pi (p)$ by contracting in Eq.\ (\ref{eq:CF1}) the relevant heavy and light
quark fields. As a result, we get%
\begin{eqnarray}
\Pi ^{\mathrm{OPE}}(p) &=&i\int d^{4}xe^{ip x}\epsilon \widetilde{\epsilon }%
\epsilon ^{\prime }\widetilde{\epsilon }^{\prime }\mathrm{Tr}\left[ \gamma
_{5}\widetilde{S}_{b}^{bb^{\prime }}(x)\gamma _{5}S_{s}^{cc^{\prime }}(x)%
\right]  \notag \\
&&\times \mathrm{Tr}\left[ \gamma _{5}\widetilde{S}_{d}^{e^{\prime
}e}(-x)\gamma _{5}S_{u}^{d^{\prime }d}(-x)\right] ,  \label{eq:OPE1}
\end{eqnarray}%
where $S_{b}(x)$ and $S_{u(d,s)}(x)$ are the heavy $b$- and light $u(d,s)$%
-quark propagators, respectively. Here we also use the shorthand notation
\begin{equation}
\widetilde{S}(x)=CS^{T}(x)C.  \label{eq:Notation}
\end{equation}%
The explicit expressions of the heavy and light quark propagators can be
found in Ref.\ \cite{Sundu:2018uyi}, for example.They contain the
perturbative and nonperturbative components: the latter depends on vacuum
expectation values of various quark, gluon, and mixed operators which
generate dependence of $\Pi ^{\mathrm{OPE}}(p)$ on the nonperturbative
quantities.

The sum rule can be extracted by equating the amplitudes $\Pi ^{\mathrm{Phys}%
}(p^{2})$ and $\Pi ^{\mathrm{OPE}}(p^{2})$, which is the first stage of the
analysis. Afterwards, we apply the Borel transformation to both sides of
this equality, this is required to suppress contributions of higher
resonances and continuum states. Next, we carry out the continuum
subtraction by invoking the assumption on the quark-hadron duality. The
obtained equality can be used to derive sum rules for $m_{T}$ and $f_{T}$,
but there is a necessity to find the second expression. As usual, it is
obtained from the first equality by applying the operator $d/d(-1/M^{2})$.
We also follow this recipe and find
\begin{equation}
m_{T}^{2}=\frac{\int_{\mathcal{M}^{2}}^{s_{0}}dss\rho ^{\mathrm{OPE}%
}(s)e^{-s/M^{2}}}{\int_{\mathcal{M}^{2}}^{s_{0}}ds\rho ^{\mathrm{OPE}%
}(s)e^{-s/M^{2}}},  \label{eq:Mass}
\end{equation}%
and%
\begin{equation}
f_{T}^{2}=\frac{1}{m_{T}^{2}}\int_{\mathcal{M}^{2}}^{s_{0}}ds\rho ^{\mathrm{%
OPE}}(s)e^{(m_{T}^{2}-s)/M^{2}},  \label{eq:Coupl}
\end{equation}%
where $\mathcal{M}=m_{b}+m_{s}$. In Eqs.\ (\ref{eq:Mass}) and (\ref{eq:Coupl}%
) $\rho ^{\mathrm{OPE}}(s)$ is the two-point spectral density, which is
proportional to the imaginary part of the correlation function $\Pi ^{%
\mathrm{OPE}}(p)$. It is seen also that the obtained sum rules have acquired
a dependence on the auxiliary parameters $M^{2}$ and $s_{0}$. The first of
them is the Borel parameter introduced during the corresponding
transformation. The $s_{0}$ is the continuum threshold parameter that
separates the ground-state and continuum contributions to $\Pi ^{\mathrm{OPE}%
}(p^{2})$ from one another.

Apart from $M^{2}$ and $s_{0}$, which are specific for each considering
problem, Eqs.\ (\ref{eq:Mass}) and (\ref{eq:Coupl}) contain vacuum
condensates%
\begin{eqnarray}
&&\langle \bar{q}q\rangle =-(0.24\pm 0.01)^{3}\ \mathrm{GeV}^{3},\ \langle
\bar{s}s\rangle =0.8\ \langle \bar{q}q\rangle ,  \notag \\
&&m_{0}^{2}=(0.8\pm 0.1)\ \mathrm{GeV}^{2},\ \langle \overline{q}g_{s}\sigma
Gq\rangle =m_{0}^{2}\langle \overline{q}q\rangle ,  \notag \\
&&\langle \overline{s}g_{s}\sigma Gs\rangle =m_{0}^{2}\langle \bar{s}%
s\rangle ,  \notag \\
&&\langle \frac{\alpha _{s}G^{2}}{\pi }\rangle =(0.012\pm 0.004)\,\mathrm{GeV%
}^{4},  \notag \\
&&\langle g_{s}^{3}G^{3}\rangle =(0.57\pm 0.29)~\mathrm{GeV}^{6}.
\label{eq:Parameters}
\end{eqnarray}%
There is also a dependence on the $b$ and $s$-quark masses, for which we use
$m_{b}=4.18_{-0.03}^{+0.04}\ \mathrm{GeV}$ and $m_{s}=96_{-4}^{+8}\ \mathrm{%
MeV}$, respectively.

In numerical computations we vary the auxiliary parameters $M^{2}$ and $%
s_{0} $ within the ranges
\begin{equation}
M^{2}\in \lbrack 3.4,\ 4.8]\ \mathrm{GeV}^{2},\ s_{0}\in \lbrack 35,\ 37]\
\mathrm{GeV}^{2}.  \label{eq:Wind1}
\end{equation}%
These windows satisfy all requirements imposed on $M^{2}$ and $s_{0}$.
Namely, the pole contribution
\begin{equation}
\mathrm{PC}=\frac{\Pi (M^{2},\ s_{0})}{\Pi (M^{2},\ \infty )},  \label{eq:PC}
\end{equation}%
where $\Pi (M^{2},\ s_{0})$ is the Borel-transformed and subtracted
invariant amplitude $\Pi ^{\mathrm{OPE}}(p^{2})$, at $M^{2}=\ 4.8\ \mathrm{%
GeV}^{2}$ is $0.18$, whereas at $M^{2}=3.4~\mathrm{GeV}^{2}$ it amounts to $%
0.63$. These two values of $M^{2}$ determine the boundaries of the region
within of which the Borel parameter can be varied. The lower limit of $M^{2}$
should meet also the very important constraint: \ the minimum of $M^{2}$ has
to ensure the convergence of the operator product expansion (OPE). This
restriction is quantified by the ratio%
\begin{equation}
R(M^{2})=\frac{\Pi ^{\mathrm{DimN}}(M^{2},\ s_{0})}{\Pi (M^{2},\ s_{0})}.
\label{eq:Convergence}
\end{equation}%
Here $\Pi ^{\mathrm{DimN}}(M^{2},\ s_{0})$ denotes a contribution to the
correlation function of the last term (or a sum of last few terms) in OPE.
Numerical analysis shows that for $\mathrm{DimN}=\mathrm{Dim(8+9+10)}$ this
ratio is $R(3.4~\mathrm{GeV}^{2})=0.013$, which guarantees the convergence
of the sum rules. Additionally, at minimal value of the Borel parameter the
perturbative term gives $62\%$ of the total result exceeding considerably
the nonperturbative contributions.

Because $M^{2}$ and $s_{0}$ are the auxiliary parameters, the mass $m_{T}$
and coupling $f_{T}$ should not depend on them. But in real calculations
there is a residual dependence of $m_{T}$ and $f_{T}$ on these parameters.
Therefore, the choice of $M^{2}$ and $s_{0}$ should minimize these
non-physical effects. The working windows for the parameters $M^{2}$ and $%
s_{0}$ given by Eq.\ (\ref{eq:Wind1}) satisfy these conditions as well. To
visualize effects of $M^{2}$ and $s_{0}$ on the mass $m_{T}$ and coupling $%
f_{T}$ we depict them in Figs.\ \ref{fig:MassT} and \ref{fig:CouplT} as
functions of these parameters. As is seen both $m_{T}$ and $f_{T}$ depend on
$M^{2}$ and $s_{0}$, which is a main source of the theoretical uncertainties
inherent to the sum rule computations. For the mass $m_{T}$ these
uncertainties are small $\pm 3\%$, because the relevant sum rule (\ref%
{eq:Mass}) is the ratio of the integrals of the functions $s\rho ^{\mathrm{%
OPE}}(s)$ and $\rho ^{\mathrm{OPE}}(s)$ which smooths these effects, but
even in the case of the coupling $f_{T}$ they do not exceed $\pm 24\%$ part
of the central value.
\begin{widetext}

\begin{figure}[h!]
\begin{center}
\includegraphics[totalheight=6cm,width=8cm]{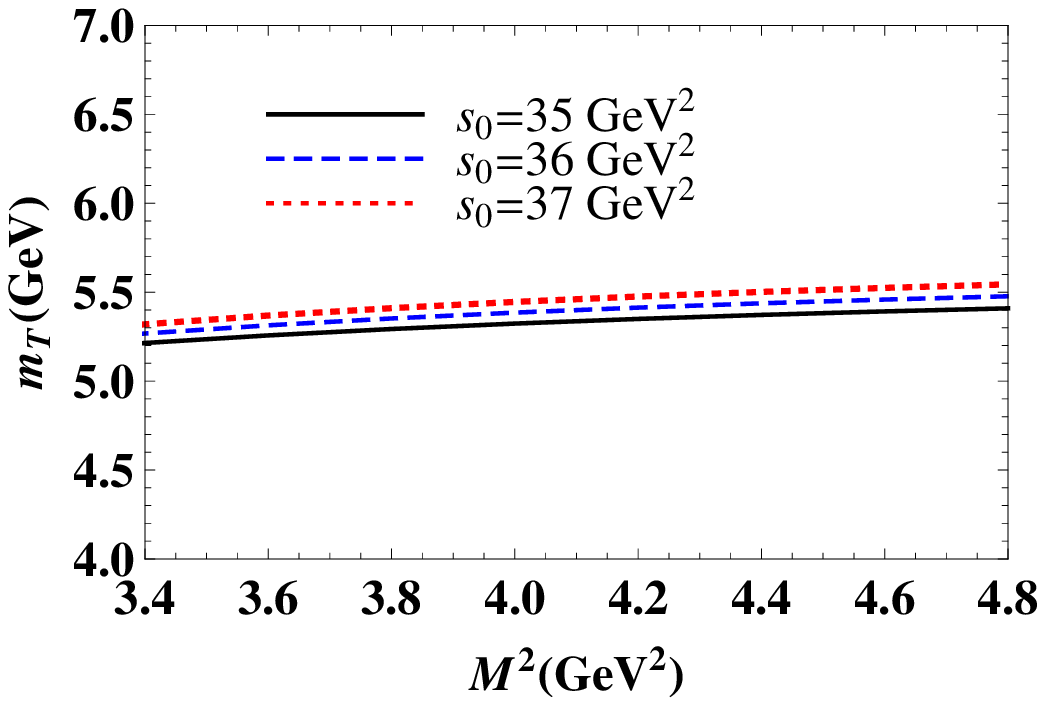}\,\, %
\includegraphics[totalheight=6cm,width=8cm]{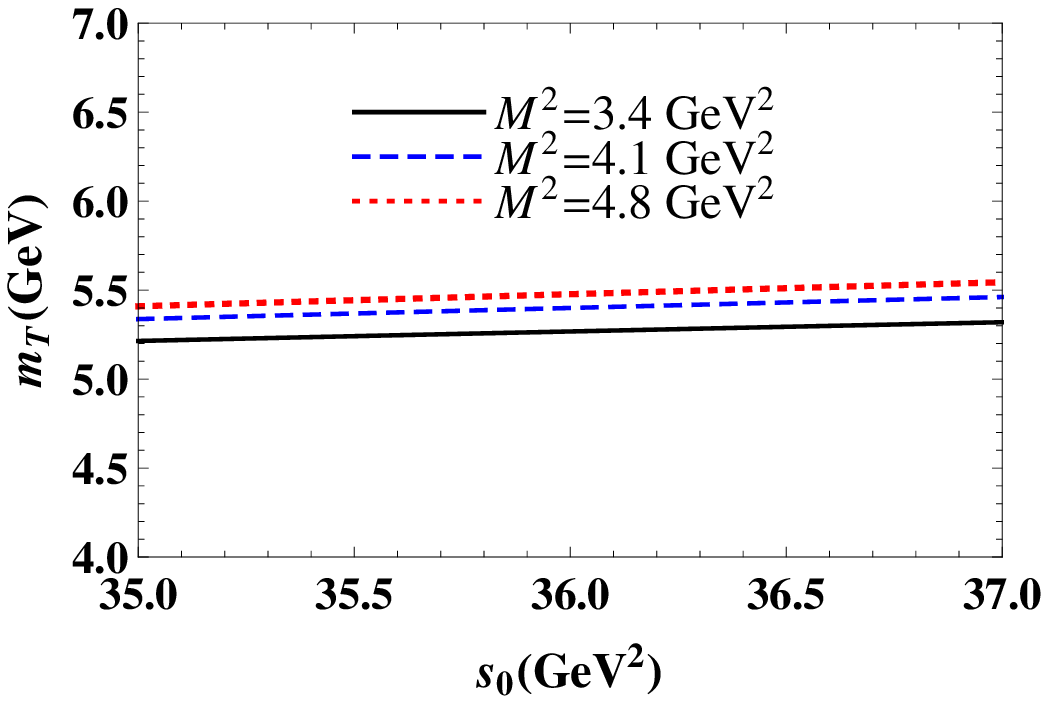}
\end{center}
\caption{ The mass of the tetraquark $T$ as a function of the Borel parameter
$M^2$ at fixed $s_0$ (left panel) and as a function of the continuum threshold
$s_0$ at fixed $M^2$ (right panel).}
\label{fig:MassT}
\end{figure}
\begin{figure}[h!]
\begin{center}
\includegraphics[totalheight=6cm,width=8cm]{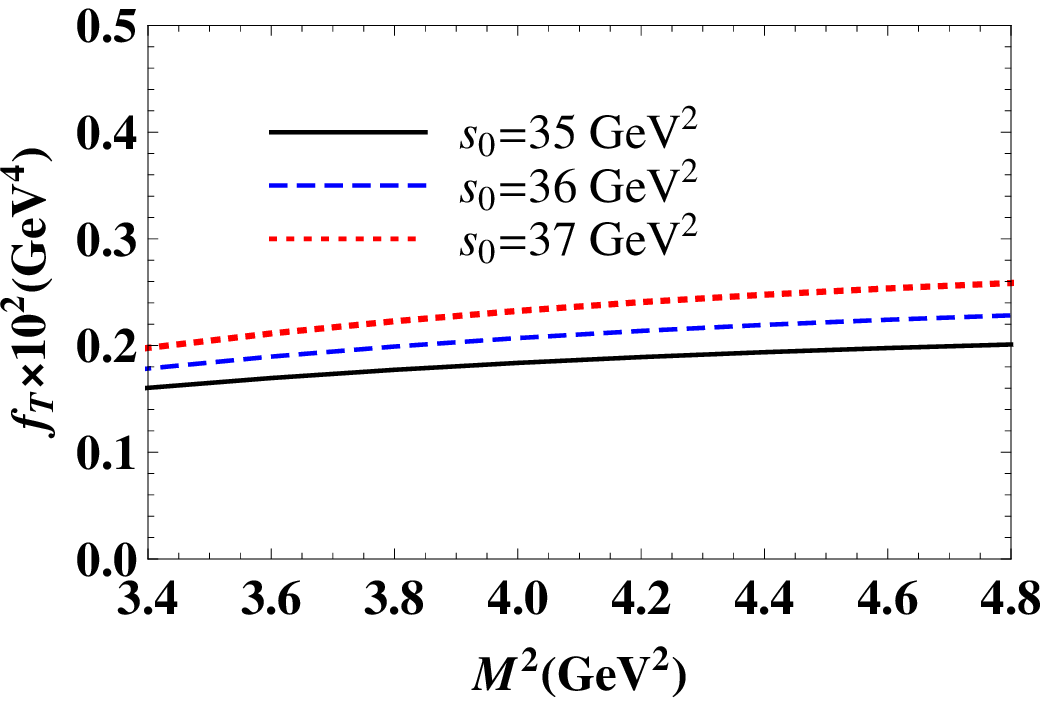}\,\, %
\includegraphics[totalheight=6cm,width=8cm]{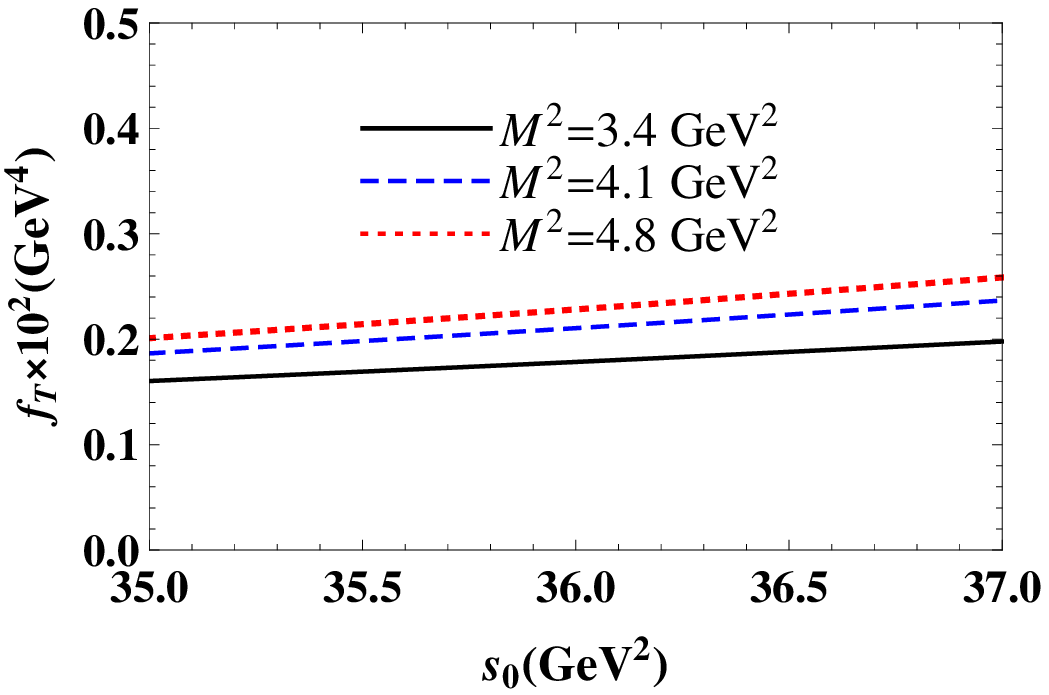}
\end{center}
\caption{ The same as in Fig.\ 1, but for the coupling $f_{T}$ of the state $T$.}
\label{fig:CouplT}
\end{figure}

\end{widetext}Our calculations for the spectroscopic parameters of the
tetraquark $T$ lead to the following results:
\begin{eqnarray}
m_{T} &=&(5380~\pm 170)~\mathrm{MeV},  \notag \\
f_{T} &=&(2.1\pm 0.5)\times 10^{-3}\ \mathrm{GeV}^{4}.  \label{eq:CMass1}
\end{eqnarray}%
The mass of the tetraquarks $T$ allows us to see whether this four-quark
meson is strong-interaction stable or not. As we have emphasized above, $T$
contains the same quark species like the resonance $X(5568)$, but differs
from it by an internal organization. The resonance $X(5568)$ with the
content $su\overline{b}\overline{d}$ was originally studied in our work \cite%
{Agaev:2016mjb}. It is a scalar particle, but has the heavy
diquark-antidiquark structure. The mass of the resonance $X(5568)$ evaluated
there%
\begin{equation}
m_{X}=(5584~\pm 137)~\mathrm{MeV}  \label{eq:Mass4}
\end{equation}%
is higher than the mass of the tetraquark $T$; structures with a heavy
diquark and a light antidiquark seem are more compact than ones composed of
a pair of heavy diquark and antidiquark. The resonance $X(5568)$ is unstable
against the strong interactions and decays to the conventional mesons $%
B_{s}^{0}\pi ^{+}$. It is clear that $T$ cannot decay to such final states,
but its quark content and quantum numbers does not forbid $S$-wave decays to
$\overline{B^{0}}K^{-}/\overline{K^{0}}B^{-}$ mesons, thresholds of which $%
5774/5777~\mathrm{MeV}$ however, are above the mass $m_{T}$. Thresholds for $%
P$-wave decays of the scalar tetraquark $bs\overline{u}\overline{d}$ are
higher than $m_{T}$ as well. The possible electromagnetic decay $%
T\rightarrow B^{-}K_{1}\gamma $ may be realized only if $m_{T}\geq 6552\
\mathrm{MeV}$, which is not the case. Therefore, transformation of the
tetraquark $T$ to ordinary mesons runs only due to its weak decays.


\section{Semileptonic decays $Z_{bc}^{0}\rightarrow Te^{+}\protect\nu_{e}$
and $Z_{bc}^{0}\rightarrow T\protect\mu ^{+}\protect\nu_{\protect\mu }$}

\label{sec:Decays}
In this section we explore the semileptonic decays $Z_{bc}^{0}\rightarrow
Te^{+}\nu _{e}$ and $Z_{bc}^{0}\rightarrow T\mu ^{+}\nu _{\mu }$ of the
scalar four-quark meson $Z_{bc}^{0}$. The spectroscopic parameters of $%
Z_{bc}^{0}$ evaluated in Ref.\ \cite{Agaev:2018khe}, as well as the mass and
coupling of the final-state tetraquark $T$ , obtained in the previous
section provide necessary information to calculate the differential rate and
width of these decays.

The decay $Z_{bc}^{0}\rightarrow T\overline{l}\nu _{l}$ runs through the
sequence of transformations $c\rightarrow W^{+}s$ and $W^{+}\rightarrow
\overline{l}\nu _{l}$, and processes with $l=e$ and $\mu $ are kinematically
allowed ones. At the tree level the transition $c\rightarrow s$ is described
by the effective Hamiltonian
\begin{equation}
\mathcal{H}^{\mathrm{eff}}=\frac{G_{F}}{\sqrt{2}}V_{cs}\overline{s}\gamma
_{\mu }(1-\gamma _{5})c\overline{l}\gamma ^{\mu }(1-\gamma _{5})\nu _{l},
\label{eq:EffH}
\end{equation}%
where $G_{F}$ is the Fermi coupling constant and $V_{cs}$ is the CKM matrix
element. Sandwiching $\mathcal{H}^{\mathrm{eff}}$ between the initial and
final tetraquarks, and factoring out the lepton fields we get the matrix
element of the current%
\begin{equation}
J_{\mu }^{\mathrm{tr}}=\overline{s}\gamma _{\mu }(1-\gamma _{5})c.
\label{eq:WeakC1}
\end{equation}%
In terms of the weak form factors $G_{i}(q^{2})$ this matrix element has the
form
\begin{equation}
\langle T(p^{\prime })|J_{\mu }^{\mathrm{tr}}|Z(p)\rangle
=G_{1}(q^{2})P_{\mu }+G_{2}(q^{2})q_{\mu },  \label{eq:Vertex1}
\end{equation}%
where $p$ and $p^{\prime }$ are the momenta of the tetraquarks $Z_{bc}^{0}$
and $T$, respectively. In Eq.\ (\ref{eq:Vertex1}) the form factors $%
G_{1}(q^{2})$ and $G_{2}(q^{2})$ parameterize the long-distance dynamics of
the weak transition. Here we also use $P_{\mu }=p_{\mu }^{\prime }+p_{\mu }$
and $q_{\mu }=p_{\mu }-p_{\mu }^{\prime }$. The $q_{\mu }$ is the momentum
transferred to the leptons, and evidently $q^{2}$ changes within the limits $%
m_{l}^{2}\leq q^{2}\leq (m_{Z}-m_{T})^{2}$, where $m_{l}$ is the mass of a
lepton $l$.

To derive the sum rules for the form factors $G_{i}(q^{2}),i=1,2$ we begin
from the three-point correlation function
\begin{eqnarray}
\Pi _{\mu }(p,p^{\prime }) &=&i^{2}\int d^{4}xd^{4}ye^{i(p^{\prime }y-px)}
\notag \\
&&\times \langle 0|\mathcal{T}\{J(y)J_{\mu }^{\mathrm{tr}}(0)J^{Z\dagger
}(x)\}|0\rangle ,  \label{eq:CF2}
\end{eqnarray}%
where $J(y)$ and $J^{Z}(x)$ are the interpolating currents for the states $T$
and $Z_{bc}^{0}$, respectively. The current $J(y)$ has been defined above by
Eq.\ (\ref{eq:Curr1}): for $J^{Z}(x)$ we use the expression \cite%
{Agaev:2018khe}%
\begin{eqnarray}
J^{Z}(x) &=&b_{a}^{T}(x)C\gamma _{5}c_{b}(x)\left[ \overline{u}_{a}(x)\gamma
_{5}C\overline{d}_{b}^{T}(x)\right.  \notag \\
&&\left. -\overline{u}_{b}(x)\gamma _{5}C\overline{d}_{a}^{T}(x)\right] .
\label{eq:Curr3}
\end{eqnarray}%
The current $J^{Z}(x)$ is composed of the $S$-wave diquark fields, has the
antisymmetric color structure $[\overline{3}_{c}]_{bc}\otimes \lbrack
3_{c}]_{\overline{u}\overline{d}}$ and describes the ground-state tetraquark
$Z_{bc}^{0}$.

As usual, we express the correlation function $\Pi _{\mu }(p,p^{\prime })$
in terms of the spectroscopic parameters of the involved particles, and find
the physical side of the sum rule $\Pi _{\mu }^{\mathrm{Phys}}(p,p^{\prime
}) $. \ The function $\Pi _{\mu }^{\mathrm{Phys}}(p,p^{\prime })$ can be
easily written down as%
\begin{eqnarray}
&&\Pi _{\mu }^{\mathrm{Phys}}(p,p^{\prime })=\frac{\langle 0|J|T(p^{\prime
})\rangle \langle T(p^{\prime })|J_{\mu }^{\mathrm{tr}}|Z(p)\rangle }{%
(p^{2}-m_{Z}^{2})(p^{\prime 2}-m_{T}^{2})}  \notag \\
&&\times \langle Z(p)|J^{Z\dagger }|0\rangle +\ldots ,  \label{eq:Phys2}
\end{eqnarray}%
where we take explicitly into account contribution only of the ground-state
particles, and denote by dots effects of the excited and continuum states.

The phenomenological side of the sum rules can be further simplified by
rewriting the relevant matrix elements in terms of the tetraquark's
parameters, and employing for $\langle T(p^{\prime })|J_{\mu }^{\mathrm{tr}%
}|Z(p)\rangle $ its expression through the weak transition form factors $%
G_{i}(q^{2})$. To this end, we use Eq.\ (\ref{eq:ME1}) and the matrix
element of the state $Z_{bc}^{0}$ defined by
\begin{equation}
\langle Z(p)|J^{Z\dagger}|0\rangle =f_{Z}m_{Z}.  \label{eq:ME3}
\end{equation}%
Then it is not difficult to find that
\begin{eqnarray}
\Pi _{\mu }^{\mathrm{Phys}}(p,p^{\prime }) &=&\frac{f_{T}m_{T}f_{Z}m_{Z}}{%
(p^{2}-m_{Z}^{2})(p^{\prime 2}-m_{T}^{2})}  \notag \\
&&\times \left[ G_{1}(q^{2})P_{\mu }+G_{2}(q^{2})q_{\mu }\right] .
\label{eq:Phys3}
\end{eqnarray}

We determine also $\Pi _{\mu }(p,p^{\prime })$ by employing the
interpolating currents and quark propagators, which lead to its expression
in terms of quark, gluon, and mixed vacuum condensates. In terms of the
quark-gluon degrees of freedom $\Pi _{\mu }(p,p^{\prime })$ takes the form
\begin{eqnarray}
&&\Pi _{\mu }^{\mathrm{OPE}}(p,p^{\prime })=i^{2}\int
d^{4}xd^{4}ye^{i(p^{\prime }y-px)}\epsilon \widetilde{\epsilon }\mathrm{Tr}%
\left[ \gamma _{\mu }(1-\gamma _{5})\right.  \notag \\
&&\left. \times S_{c}^{ib^{\prime }}(-x)\gamma _{5}\widetilde{S}%
_{b}^{ba^{\prime }}(y-x)\gamma _{5}S_{s}^{ci}(y)\right] \left\{ \mathrm{Tr}%
\left[ \gamma _{5}\widetilde{S}_{d}^{a^{\prime }e}(x-y)\right. \right.
\notag \\
&&\left. \left. \times \gamma _{5}S_{u}^{b^{\prime }d}(x-y)\right] -\mathrm{%
Tr}\left[ \gamma _{5}\widetilde{S}_{d}^{b^{\prime }e}(x-y)\gamma
_{5}S_{u}^{a^{\prime }d}(x-y)\right] \right\} ,  \notag \\
&&  \label{eq:CF4}
\end{eqnarray}%
where $a^{\prime }$, $b^{\prime }$ and $i$ are the color indices of the
currents $J^{Z}(x)$ and $J_{\mu }^{\mathrm{tr}}$, respectively.

We extract the sum rules for the form factors $G_{i}(q^{2})$ by equating the
invariant amplitudes corresponding to the same Lorentz structures in $\Pi
_{\mu }^{\mathrm{Phys}}(p,p^{\prime })$ and $\Pi _{\mu }^{\mathrm{OPE}%
}(p,p^{\prime })$. After that, we carry out the double Borel transformation
over the variables $p^{\prime 2}$ and $p^{2}$ necessary to suppress
contributions of the higher excited and continuum states, and finally carry
out the continuum subtraction. These manipulations yield the sum rules
\begin{eqnarray}
&&G_{i}(\mathbf{M}^{2},\ \mathbf{s}_{0},~q^{2})=\frac{1}{f_{T}m_{T}f_{Z}m_{Z}%
}\int_{(m_{b}+m_{c})^{2}}^{s_{0}}ds  \notag \\
&&\times \int_{\mathcal{M}^{2}}^{s_{0}^{\prime }}ds^{\prime }\rho
_{i}(s,s^{\prime },q^{2})e^{(m_{Z}^{2}-s)/M_{1}^{2}}e^{(m_{T}^{2}-s^{\prime
})/M_{2}^{2}}.  \label{eq:FF}
\end{eqnarray}%
Here $\mathbf{M}^{2}=(M_{1}^{2},\ M_{2}^{2})$ and $\mathbf{s}_{0}=(s_{0},\
s_{0}^{\prime })$ are the Borel and continuum threshold parameters,
respectively. It is worth noting that the set $(M_{1}^{2},s_{0})$ describes $%
Z_{bc}^{0}$, whereas $(M_{2}^{2},s_{0}^{\prime })$ corresponds to the $T$
tetraquark channel. The spectral densities $\rho _{i}(s,s^{\prime },q^{2})$
are calculated as the imaginary parts of the correlation function $\Pi _{\mu
}^{\mathrm{OPE}}(p,p^{\prime })$ with dimension-five accuracy, and contain
both the perturbative and nonperturbative contributions.

For numerical computations of $G_{i}(\mathbf{M}^{2},\ \mathbf{s}_{0},~q^{2})$
one needs to employ various parameters, values some of which are collected
in Eq.\ (\ref{eq:Parameters}). The mass and coupling of the tetraquark $%
Z_{bc}^{0}$ and $(M_{1}^{2},s_{0})$ are borrowed from Ref. \cite%
{Agaev:2018khe}, whereas for $m_{T}$ and $f_{T}$, and $(M_{2}^{2},s_{0}^{%
\prime})$ we use results of the previous section.

To obtain the width of the decay $Z_{bc}^{0}\rightarrow T\overline{l}v_{l}$\
we have to integrate the differential decay rate $d\Gamma /dq^{2}$ within
the kinematical limits $m_{l}^{2}\leq q^{2}\leq (m_{Z}-m_{T})^{2}$, whereas
the QCD sum rules lead to reliable results only for $m_{l}^{2}\leq q^{2}\leq
1.25$ $\mathrm{GeV}^{2}$. To cover all values of $q^{2}$ we replace the weak
form factors by the functions $F_{i}(q^{2})$, which at accessible for the
sum rule computations $q^{2}$ coincide with $G_{i}(q^{2})$, but can be
extrapolated to the whole integration region.

\begin{figure}[h!]
\begin{center}
\includegraphics[totalheight=6cm,width=8cm]{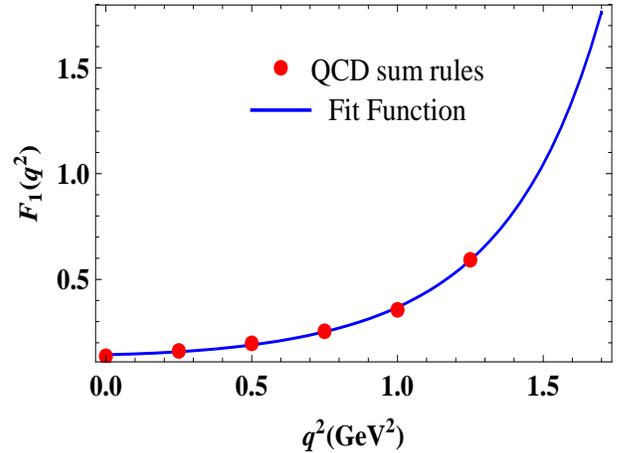}
\end{center}
\caption{The sum rule predictions for the weak form factor $G_{1}(q^{2})$
and the fit function $F_{1}(q^{2})$. }
\label{fig:FitF1}
\end{figure}

In the present work for the fit functions we utilize the analytic
expressions
\begin{equation}
F_{i}(q^{2})=f_{i}^{0}\exp \left[ c_{1}^{i}\frac{q^{2}}{m_{Z}^{2}}%
+c_{2}^{i}\left( \frac{q^{2}}{m_{Z}^{2}}\right) ^{2}\right] .
\label{eq:FFunctions}
\end{equation}%
Here, $f_{i}^{0},~c_{1}^{i}$ and $c_{2}^{i}$ are fitting parameters, values
of which are presented below%
\begin{eqnarray}
f_{1}^{0} &=&0.144,\ c_{1}^{1}=7.68,\ c_{2}^{1}=1505.10,  \notag \\
f_{2}^{0} &=&3.282,\ c_{1}^{2}=7.69,\ c_{2}^{2}=1504.40.  \label{eq:FitPar}
\end{eqnarray}%
In Fig.\ \ref{fig:FitF1}, as an example, we plot the sum rule predictions
for the form factor $G_{1}(q^{2})$ and the fit function $F_{1}(q^{2})$ : It
is seen that the fit function coincides well with the sum rule predictions
in the region $m_{l}^{2}\leq q^{2}\leq 1.25$ $\mathrm{GeV}^{2}$.

The differential rate $d\Gamma /dq^{2}$ of the semileptonic decay $%
Z_{bc}^{0}\rightarrow T\overline{l}\nu _{l}$ is given by the formula
\begin{widetext}
\begin{eqnarray}
\frac{d\Gamma }{dq^{2}} &=&\frac{C_{F}^{2}|V_{cs}|^{2}}{64\pi ^{3}m_{Z}^{3}}%
\lambda \left( m_{Z}^{2},m_{T}^{2},q^{2}\right) \left( \frac{q^{2}-m_{l}^{2}%
}{q^{2}}\right) ^{2}\left\{ (2q^{2}+m_{l}^{2})\left[ |G_{1}(q^{2})|^{2}%
\left( \frac{q^{2}}{2}-m_{Z}^{2}-m_{T}^{2}\right) -|G_{2}(q^{2})|^{2}\frac{%
q^{2}}{2}\right. \right.   \notag \\
&&\left. +(m_{T}^{2}-m_{Z}^{2})\mathrm{\func{Re}}\left[ G_{1}(q^{2})G_{2}^{%
\ast }(q^{2})\right] \right] +\frac{q^{2}+m_{l}^{2}}{q^{2}}\left[
|G_{1}(q^{2})|^{2}(m_{Z}^{2}-m_{T}^{2})^{2}+|G_{2}(q^{2})|^{2}q^{4}+2\mathrm{%
\func{Re}}\left[ G_{1}(q^{2})G_{2}^{\ast }(q^{2})\right] \right.   \notag \\
&&\left. \left. \times (m_{Z}^{2}-m_{T}^{2})q^{2}\right] \right\} ,
\label{eq:Drate}
\end{eqnarray}%
where%
\begin{equation}
\lambda \left( m_{Z}^{2},m_{T}^{2},q^{2}\right) =\left[
m_{Z}^{4}+m_{T}^{4}+q^{4}-2\left(
m_{Z}^{2}m_{T}^{2}+m_{Z}^{2}q^{2}+m_{T}^{2}q^{2}\right) \right] ^{1/2}.
\label{eq:Lambda}
\end{equation}
\end{widetext}To fulfil the numerical computations using Eq.(\ref{eq:Drate})
one also needs the Fermi coupling constant $G_{F}=1.16637\times 10^{-5}~%
\mathrm{GeV}^{-2}$ and CKM matrix element $|V_{cs}|=0.997\pm 0.017$.
Obtained results for the width of semileptonic decays $Z_{bc}^{0}\rightarrow
T\overline{l}\nu _{l}$ ($l=e,~\mu $) read
\begin{eqnarray}
\Gamma \left( Z_{bc}^{0}\rightarrow Te^{+}\nu _{e}\right)  &=&\left( 1.19\pm
0.26\right) \times 10^{-11}\ \mathrm{MeV},  \notag \\
\Gamma \left( Z_{bc}^{0}\rightarrow T\mu ^{+}\nu _{\mu }\right)  &=&\left(
1.18\pm 0.25\right) \times 10^{-11}\ \mathrm{MeV}.  \notag \\
&&  \label{eq:SemilDW}
\end{eqnarray}%
These results are important part of the information to evaluate the full
width and mean lifetime of the tetraquark $Z_{bc}^{0}$, and estimate
branching ratios of its weak decay channels.

\section{ Nonleptonic two-body decays $Z_{bc}^{0}\rightarrow T\protect\pi %
^{+}$ and $Z_{bc}^{0}\rightarrow TK^{+}$}

\label{sec:Decays2}

The nonleptonic two-body  decays $Z_{bc}^{0}\rightarrow T\pi ^{+}$ and $%
Z_{bc}^{0}\rightarrow TK^{+}$ of the tetraquark $Z_{bc}^{0}$ can be
considered in the context of the QCD factorization approach, which allows us
to calculate the amplitudes and widths of these processes. This method was
successfully applied to study two-body weak decays of the conventional
mesons \cite{Beneke:1999br,Beneke:2000ry}, and is used here to investigate
two-body decays of the tetraquark $Z_{bc}^{0}$, when one of the final
particles is an exotic meson.

At the quark level, the effective Hamiltonian for the decay $%
Z_{bc}^{0}\rightarrow T\pi ^{+}$ is given by the expression
\begin{equation}
\widetilde{\mathcal{H}}^{\mathrm{eff}}=\frac{G_{F}}{\sqrt{2}}%
V_{cs}V_{ud}^{\ast }\left[ c_{1}(\mu )Q_{1}+c_{2}(\mu )Q_{2}\right] ,
\label{eq:EffHam}
\end{equation}%
where%
\begin{eqnarray}
Q_{1} &=&\left( \overline{u}_{i}d_{i}\right) _{\mathrm{V-A}}\left( \overline{%
s}_{j}c_{j}\right) _{\mathrm{V-A}},  \notag \\
Q_{2} &=&\left( \overline{u}_{i}d_{j}\right) _{\mathrm{V-A}}\left( \overline{%
s}_{j}c_{i}\right) _{\mathrm{V-A}},  \label{eq:Operators}
\end{eqnarray}%
and $i$ , $j$ are the color indices. Here $c_{1}(\mu )$ and $c_{2}(\mu )$
are the short-distance Wilson coefficients evaluated at the scale $\mu $ at
which the factorization is assumed to be correct. The shorthand notation $%
\left( \overline{q}_{1}q_{2}\right) _{\mathrm{V-A}}$ in Eq.\ (\ref%
{eq:Operators}) means
\begin{equation}
\left( \overline{q}_{1}q_{2}\right) _{\mathrm{V-A}}=\overline{q}_{1}\gamma
_{\mu }(1-\gamma _{5})q_{2}.  \label{eq:Not}
\end{equation}%
The amplitude of this decay can be written down in the following factorized
form%
\begin{eqnarray}
\mathcal{A} &=&\frac{G_{F}}{\sqrt{2}}V_{cs}V_{ud}^{\ast }a_{1}(\mu )\langle
\pi ^{+}(q)|\left( \overline{u}_{i}d_{i}\right) _{\mathrm{V-A}}|0\rangle
\notag \\
&&\times \langle T(p^{\prime })|\left( \overline{s}_{j}c_{j}\right) _{%
\mathrm{V-A}}|Z(p)\rangle ,  \label{eq:Amplitude}
\end{eqnarray}%
where
\begin{equation}
a_{1}(\mu )=c_{1}(\mu )+\frac{1}{N_{c}}c_{2}(\mu ),
\end{equation}%
with $N_{c}$ being the number of quark colors. The amplitude $\mathcal{A}$
corresponds to the process in which the pion $\pi ^{+}$ is generated
directly from the color-singlet current $\left( \overline{u}_{i}d_{i}\right)
_{\mathrm{V-A}}$. The matrix element $\langle T(p^{\prime })|\left(
\overline{s}_{j}c_{j}\right) _{\mathrm{V-A}}|Z(p)$ has been defined above in
Eq.\ (\ref{eq:Vertex1}), whereas the matrix element of the pion in given by
the expression%
\begin{equation}
\langle \pi ^{+}|\left( \overline{u}_{i}d_{i}\right) _{\mathrm{V-A}%
}|0\rangle =if_{\pi }q_{\mu }.  \label{eq:ME4}
\end{equation}%
and is determined by its decay constant $f_{\pi }$.

Then, it is not difficult to see that $\mathcal{A}$ takes the form%
\begin{eqnarray}
\mathcal{A} &=&i\frac{G_{F}}{\sqrt{2}}f_{\pi }V_{cs}V_{ud}^{\ast }a_{1}(\mu )
\notag \\
&&\times \left[ G_{1}(q^{2})Pq+G_{2}(q^{2})q^{2}\right] .
\label{eq:Amplitude2}
\end{eqnarray}%
The width of the decay $Z_{bc}^{0}\rightarrow T\pi ^{+}$ is equal to:%
\begin{eqnarray}
&&\Gamma \left( Z_{bc}^{0}\rightarrow T\pi ^{+}\right) =\frac{%
G_{F}^{2}f_{\pi }^{2}}{32\pi m_{Z}^{3}}|V_{cs}|^{2}|V_{ud}|^{2}a_{1}^{2}(\mu
)  \notag \\
&&\times \lambda \left[ |G_{1}(m_{\pi
}^{2})|^{2}(m_{Z}^{2}-m_{T}^{2})^{2}+|G_{2}(m_{\pi }^{2})|^{2}m_{\pi
}^{4}\right.  \notag \\
&&\left. +2\func{Re}\left[ G_{1}(m_{\pi }^{2})G_{2}^{\ast }(m_{\pi }^{2})%
\right] (m_{Z}^{2}-m_{T}^{2})m_{\pi }^{2}\right] ,  \label{eq:NonLDW}
\end{eqnarray}%
where $\lambda =\lambda (m_{Z}^{2},m_{T}^{2},m_{\pi }^{2})$ is the function
given by Eq. (\ref{eq:Lambda}). The similar analysis is valid for the second
decay $Z_{bc}^{0}\rightarrow TK^{+}$, as well: relevant formulas can by
obtained by replacements $V_{ud}\rightarrow V_{us}$, $f_{\pi }\rightarrow
f_{K}$, and $m_{\pi }\rightarrow m_{K}$.

Numerical computations can be carried out after fixing the spectroscopic
parameters of the light mesons $\pi ^{+}$ and $K^{+}$. In calculations we
use $m_{\pi }=139.570\ \mathrm{MeV}$, $f_{\pi }=131\ \mathrm{MeV}$, and $%
m_{K}=\left( 493.677\pm 0.016\right) \ \mathrm{MeV}$, $f_{K}=(155.72\pm
0.51)\ \mathrm{MeV}$, respectively. The weak form factors $G_{1}(q^{2})$ and
$G_{2}(q^{2})$, which are main ingredients of $\Gamma \left(
Z_{bc}^{0}\rightarrow T\pi ^{+}(K^{+})\right) $, have been obtained in the
previous section. For CKM matrix elements we use $|V_{ud}|=0.974$ and $%
|V_{us}|=0.224$. The Wilson coefficients at the factorization scale $\mu
=m_{c}$ are borrowed from Ref.\ \cite{Colangelo:2001cv}%
\begin{equation}
c_{1}(m_{c})=1.263,\ c_{2}(m_{c})=-0.513.  \label{eq:WCoeff}
\end{equation}

For the decay $Z_{bc}^{0}\rightarrow T\pi ^{+}$, our calculations lead to
the result
\begin{eqnarray}
&&\Gamma \left( Z_{bc}^{0}\rightarrow T\pi ^{+}\right) =\left( 7.05\pm
1.52\right) \times 10^{-12}\ \mathrm{MeV},  \notag \\
&&  \label{eq:NLDW1}
\end{eqnarray}%
which is smaller than widths of the semileptonic decays, but nevertheless is
comparable with them. For the second process $Z_{bc}^{0}\rightarrow TK^{+}$
we get
\begin{eqnarray}
&&\Gamma \left( Z_{bc}^{0}\rightarrow TK^{+}\right) =\left( 1.02\pm
0.21\right) \times 10^{-12}~\mathrm{MeV}.  \notag \\
&&  \label{eq:NLDW2}
\end{eqnarray}%
It is not difficult to see that effect of this decay to formation of the
full width of the tetraquark $Z_{bc}^{0}$ is very small. The partial widths
of the nonleptonic two-body decays obtained in this section will be used
below to find the full width of $Z_{bc}^{0}$.


\section{Analysis and concluding remarks}

\label{sec:Analysis}
The partial widths of the dominant semileptonic and two nonleptonic decay
modes of $Z_{bc}^{0}$ allow us to evaluate its full width and mean lifetime
\begin{eqnarray}
\Gamma _{\mathrm{full}} &=&(3.18\pm 0.39)\times 10^{-11}~\mathrm{MeV},\
\notag \\
\tau &=&2.07_{-0.23}^{+0.29}\times 10^{-11}~\mathrm{s}.  \label{eq:FR}
\end{eqnarray}%
As is seen, the scalar tetraquark $Z_{bc}^{0}$ is narrower than the master
particle $T_{bb;\overline{u}\overline{d}}^{-}$, and its mean lifetime $%
20.7_{-2.3}^{+2.9}~\mathrm{ps}$ is considerably longer that the same
parameter for $T_{bb;\overline{u}\overline{d}}^{-}$.

The weak decays of $Z_{bc}^{0}$ occur via the following channels:

i) $Z_{bc}^{0}\rightarrow Te^{+}\nu_{e}$,

ii) $Z_{bc}^{0}\rightarrow T\mu ^{+}\nu _{\mu }$,

iii) $Z_{bc}^{0}\rightarrow T\pi ^{+}$,

and

iv) $Z_{bc}^{0}\rightarrow TK^{+}$.

All of them leads to appearance of the strong- and
electromagnetic-interaction stable tetraquark $T\equiv T_{bs;\overline{u}%
\overline{d}}^{-}$ that at next stages of the process dissociates weakly.
The branching ratio for production, for example, of the final state $%
Te^{+}\nu _{e}$ is given by
\begin{equation}
\mathcal{BR}(Z_{bc}^{0}\rightarrow Te^{+}\nu _{e})=\Gamma \left(
Z_{bc}^{0}\rightarrow Te^{+}\nu _{e}\right) /\Gamma _{\mathrm{full}}.
\label{eq:BR1}
\end{equation}%
It is not difficult to find that
\begin{eqnarray}
\mathcal{BR}(Z_{bc}^{0} &\rightarrow &Te^{+}\nu _{e})\simeq 0.38,~\mathcal{BR%
}(\ Z_{bc}^{0}\rightarrow T\mu ^{+}\nu _{\mu })\simeq 0.37  \notag \\
\mathcal{BR}(Z_{bc}^{0} &\rightarrow &T\pi ^{+})\simeq 0.22,~\mathcal{BR}(\
Z_{bc}^{0}\rightarrow TK^{+})\simeq 0.03.  \notag \\
&&  \label{eq:BR2}
\end{eqnarray}%
\newline

The weak decays of $T_{bb;\overline{u}\overline{d}}^{-}$ can be analyzed by
the same way. The relevant semileptonic modes at the final state contain the
tetraquark $T_{bs;\overline{u}\overline{d}}^{-}$ and two opposite sign
leptons accompanying by corresponding neutrinos $e^{-}e^{+}\nu _{e}\overline{%
\nu }_{e}$, $e^{-}\mu ^{+}\overline{\nu }_{e}\nu _{\mu }$, $e^{+}\mu ^{-}\nu
_{e}\overline{\nu }_{\mu }$, $\mu ^{+}\mu ^{-}\nu _{\mu }\overline{\nu }%
_{\mu }$, $\tau ^{-}e^{+}\nu _{e}\overline{\nu }_{\tau }$ and $\tau ^{-}\mu
^{+}\overline{\nu }_{\tau }\nu _{\mu }$. Other decay channels are formed by
the final states $Te^{-}\overline{\nu }_{e}\pi ^{+}$, $Te^{-}\overline{\nu }%
_{e}K^{+}$, \ $T\mu ^{-}\overline{\nu }_{\mu }\pi ^{+}$, $T\mu ^{-}\overline{%
\nu }_{\mu }K^{+}$, $T\tau ^{-}\overline{v}_{\tau }\pi ^{+}$, and $T\tau ^{-}%
\overline{v}_{\tau }K^{+}$. The branching ratios of these channels can be
found using the fact, that $\mathcal{BR}(T_{bb;\overline{u}\overline{d}%
}^{-}\rightarrow Z_{bc}^{0}e^{-}\overline{\nu }_{e})\simeq \mathcal{BR}%
(T_{bb;\overline{u}\overline{d}}^{-}\rightarrow Z_{bc}^{0}\mu ^{-}\overline{%
\nu }_{\mu })=0.37$ and $\mathcal{BR}(T_{bb;\overline{u}\overline{d}%
}^{-}\rightarrow Z_{bc}^{0}\tau ^{-}\overline{v}_{\tau })=0.26$ (see, Ref.
\cite{Agaev:2018khe}). For some of decay modes we get:
\begin{eqnarray}
\mathcal{BR}(T_{bb;\overline{u}\overline{d}}^{-} &\rightarrow
&Te^{-}e^{+}\nu _{e}\overline{\nu }_{e})\simeq 0.141,  \notag \\
\mathcal{BR}(T_{bb;\overline{u}\overline{d}}^{-} &\rightarrow &T\mu ^{+}\mu
^{-}\nu _{\mu }\overline{\nu }_{\mu })\simeq 0.137,  \notag \\
\mathcal{BR}(T_{bb;\overline{u}\overline{d}}^{-} &\rightarrow &T\tau
^{-}e^{+}\nu _{e}\overline{\nu }_{\tau })\simeq 0.099,  \notag \\
\mathcal{BR}(T_{bb;\overline{u}\overline{d}}^{-} &\rightarrow &Te^{-}%
\overline{\nu }_{e}\pi ^{+})\simeq 0.081,  \notag \\
\mathcal{BR}(T_{bb;\overline{u}\overline{d}}^{-} &\rightarrow &Te^{-}%
\overline{\nu }_{e}K^{+})\simeq 0.011.  \label{eq:BR3a}
\end{eqnarray}

We have explored the weak decays of the scalar tetraquark $Z_{bc}^{0}$
including its dominant semileptonic transformations to $Te^{+}\nu _{e}$ and $%
T\mu ^{+}\nu _{\mu }$, as well as the two-body nonleptonic decays $%
Z_{bc}^{0}\rightarrow T\pi ^{+}$ and $Z_{bc}^{0}\rightarrow TK^{+}$, and
estimated branching ratios of these final states. Because $Z_{bc}^{0}$ is
stable against strong and electromagnetic decays, weak modes are important
for its experimental studies: in accordance with recent analysis the
production rate of the tetraquarks with the heavy diquark $bc$ at the LHC
would be higher by two order of magnitude than four-quark mesons with $bb$
\cite{Ali:2018xfq}.

Another issue studied here is decays of the tetraquark $T_{bb;\overline{u}%
\overline{d}}^{-}$. We have analyzed its decay chains consisting of
sequential weak transformations to final states with $T$ and evaluated their
branching ratios. These calculations are important to fix processes, where
the axial-vector tetraquark $T_{bb;\overline{u}\overline{d}}^{-}$ should be
searched for.

The predictions for the width and lifetime of $Z_{bc}^{0}$, as well as for
the branching ratios (\ref{eq:BR2}) and (\ref{eq:BR3a}) should be considered
as first results for these quantities obtained using dominant weak decays of
$Z_{bc}^{0}$ and $T_{bb;\overline{u}\overline{d}}^{-}$. In fact, here we
have taken into account only processes $Z_{bc}^{0}\rightarrow Te^{+}\nu _{e}$%
,\ $Z_{bc}^{0}\rightarrow T\mu ^{+}\nu _{\mu }$, $Z_{bc}^{0}\rightarrow T\pi
^{+}$ and $Z_{bc}^{0}\rightarrow TK^{+}$, but subdominant semileptonic
decays of $Z_{bc}^{0}$ may correct these predictions. We have treated $T$ as
a scalar particle, whereas $Z_{bc}^{0}$ can decay also to exotic mesons with
another quantum numbers. By including into analysis these options one can
open up new decay modes of $Z_{bc}^{0}$, and improve predictions for the
branching ratios presented above. Finally, there are nonleptonic three-meson
decay channels, effects of which on the full width and mean lifetime of $%
Z_{bc}^{0}$ maybe sizeable. In other words, nonleading semileptonic decays
of $Z_{bc}^{0}$, its decays to a tetraquark $T$ with another quantum
numbers, and to multimeson nonleptonic final states may improve and correct
the picture described here. Detailed investigations of these problems, left
beyond the scope of the present work, are necessary to gain more precise
knowledge about properties of the exotic states $T_{bb;\bar{u}\bar{d}}^{-}$
and $Z_{bc}^{0}$.

.

\end{document}